\documentclass[aps, prb, amsmath,amssymb,reprint]{revtex4-1}

\usepackage{graphicx}				
\usepackage{bm}					
\usepackage{siunitx}
\usepackage{xspace}
\usepackage[final]{changes}
\usepackage[colorlinks=true,linkcolor=red,urlcolor=blue,citecolor=cyan]{hyperref}

\newcommand{\FIGW}{\columnwidth}

\begin{document}

\title{Tunable Finite-Sized Chains to Control Magnetic Relaxation}

\author{Paul D. Ekstrand}
\author{Daniel J. Javier}
\author{Thomas Gredig} \email{tgredig@csulb.edu}

\affiliation{ 
Department of Physics and Astronomy,
California State University Long Beach, CA 90840-9505, U.S.A.
}

\date{\today}

\begin{abstract}

The magnetic dynamics of low-dimensional iron ion chains have been studied with regards to the tunable finite-sized chain length using iron phthalocyanine thin films. The deposition temperature varies the diffusion length during thin film growth by limiting the average crystal size in the range from \SI{40}{nm} to \SI{110}{nm}. Using a method common for single chain magnets, the magnetic relaxation time for each chain length is determined from temporal remanence data and fit to a stretched exponential form in the temperature range below \SI{5}{K}, the onset for magnetic hysteresis. A temperature-independent master curve is generated by scaling the remanence by its relaxation time to fit the energy barrier for spin reversal, and the single spin relaxation time. The energy barrier of \SI{95}{K} is found to be independent of the chain length. In contrast, the single spin relaxation time increases with longer chains from under 1~ps to 800~ps.  We show that thin films provide the nano-architecture to control magnetic relaxation and a testbed to study finite-size effects in low-dimensional magnetic systems.

\end{abstract}


\maketitle


Low-dimensional magnetic systems, including single molecular magnets and single chain magnets, not only provide insight into fundamental kinetics of magnetism, but also act as seeds for novel systems with optical properties not accessible in conventional magnets.  Single chain magnets consist of strongly coupled magnetic ions along a chain with negligible inter-chain interactions.\cite{coulon_single-chain_2006,christou_single-molecule_2000} Magnetic hysteresis loops of such low-dimensional systems are not the result of an equilibrium state, instead represent a dynamic system with slow relaxation. The time-dependence of the magnetic system with special anisotropy can be understood in the theoretical framework of the Glauber-Ising model.\cite{glauber_timedependent_1963,coulon_glauber_2004} 

Despite the rapid progress of theoretical models for magnetic dynamics in low-dimensional systems, experimental realizations of single chain magnets have emerged only recently and are relatively few.\cite{caneschi_cobaltii-nitronyl_2001,caneschi_glauber_2002,ferbinteanu_single-chain_2005,zhang_single-chain_2013,coulon_single-chain_2014,vindigni_fast_2005,liu_light-induced_2013,wei_single-chain_2016,gatteschi_single-chain_2014} The intra-chain interactions $J'$ should be several orders of magnitude larger than the inter-chain interactions $J''$ to be considered a single chain magnet. The relaxation phenomenon has been studied  using ac susceptibility mostly in powder specimen or small crystals.\cite{caneschi_molecular_1999,bogani_molecular_2008} Experimentally, the relaxation time $\tau$ can be extracted from the frequency-dependent peak positions obtained from ac susceptibility. Additionally, dc measurements of the remanent magnetization probe the slower end of relaxation times. A link between the two measurement techniques has been demonstrated in heterometallic chains of Mn(III)-Ni(II) crystals.\cite{clerac_evidence_2002} Interpreting the relaxation times, the kinetic Ising model by Glauber provides context for justification of the Arrhenius law behavior taking the form $\tau = \tau_0 \exp(\Delta_A/k_B T)$, where $\tau_0$ is the single spin relaxation time.

The Hamiltonian in this kinetic model also contains the length $L$ of the finite-sized chain.\cite{leal_da_silva_critical_1995,vindigni_finite_2004,vindigni_finite-sized_2005,pini_finite-size_2011} Tuning this one-dimensional quantity has been a challenge.  Given impurities of complex molecules and structural point defects, the chain is randomly interrupted. In order to simulate the effect of the chain length, non-magnetic impurities have been added intentionally to a crystal.\cite{bogani_finite-size_2004} In that work, the single spin relaxation time $\tau_0$ is reduced after adding non-magnetic Zn impurities from \SI{35}{ps} in pure material to \SI{1}{ps} in the doped samples.  Controlled experiments of the finite-chain size, however, are by and large missing.  Unlike powder and crystals, thin films are ideal for building magnetic nano-structures\cite{gruber_exchange_2015} and tuning properties through substrate choice and deposition parameters. This provides a means to directly control the chain length during a careful deposition of a thin film. 

One well-studied compound for building magnetic nano-structures is the metallo-phthalocyanine molecule,  a small molecule with a 3d-metal ion center connected to pyridine ligands. Amongst the organometallics, the synthesis, structure, and growth of phthalocyanines has been extensively studied with applications for photovoltaic cells and gas sensors.\cite{yang_ultrathin_2007} 

In the seminal work on $\alpha$-phase Fe(II)-phthalocyanine (FePc) powder, relaxation behavior was noticed.\cite{evangelisti_magnetic_2002} Two hysteresis loops measured at two different sweeping speeds produced different curves. In powder specimen, the intra-chain interaction is fit from high-temperature susceptibility data to be $J' \sim$ 15 K/$k_B$ for spin $S=1$. Using Oguchi's method,\cite{oguchi_theory_1963} the inter-chain interaction was estimated to be $J'' < 10^{-3} J'$,  demonstrating quasi one-dimensional character of FePc thin films.

Similar to the powder specimen, thin film samples of FePc also show interesting magnetic properties. In particular, FePc thin films have non-equilibrium magnetic hysteresis loops below \SI{5}{K}.\cite{gredig_control_2010,bartolome_highly_2010,gredig_substrate-controlled_2012}  The coercivity in films of fixed thickness strongly depends on the deposition temperature, the measurement speed, and measurement temperature.\cite{gredig_height-height_2013}  XMCD data of FePc/Au thin films reveals an $xy$ anisotropy in the plane of the molecule.\cite{bartolome_highly_2010}   In the following, the systematic details of magnetic relaxation in FePc thin films are studied and presented as a paradigm for tunable finite-sized nano-structures to probe magnetic dynamics in low-dimensional systems.

During self-assembly of the molecules in thermal evaporation, iron ion chains form along the b-axis of the molecule. Chains are separated laterally by $\sim 1.3$~nm through a carbon matrix.  The crystal size of thin films varies strongly with deposition temperature and provides an artificial cut-off point for the average length $L$.  A quantitative study provides data on the precise distributions of these chains.\cite{gentry_asymmetric_2009} The grain size distribution is fundamentally categorized into two types, the low temperature ($T<200$~\si{\degreeCelsius}) regime and the high-temperature regime that yields larger grains, but also more roughness and pin holes due to Ehrlich-Schwoebel growth. The thin film grains are asymmetric, similar to the single-crystals in powder that have needle-like shapes. The iron ion chain follows the long axis \added{of the crystals as determined with atomic force microscopy (AFM).  The crystal length is a log-normally distributed, where the mode of the long-axis} was quantitatively extracted. It varies from $L = $~\SI{39}{nm} to nearly \SI{200}{nm}.\cite{gentry_asymmetric_2009}

\begin{figure} 
    \centering
    \includegraphics[width=\FIGW]{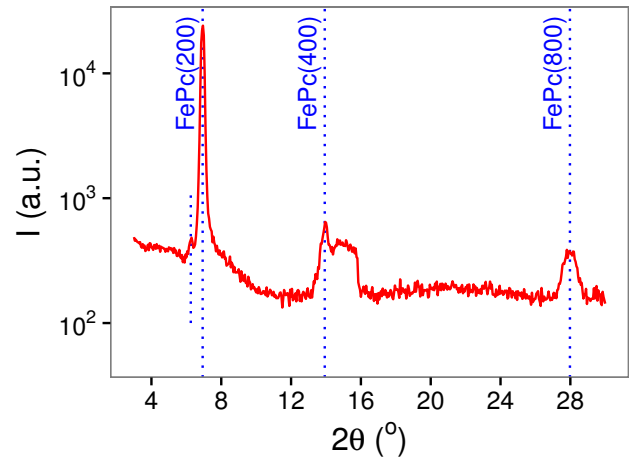}  
    \caption{The scattered intensity from x-ray diffraction of FePc/silicon sample shows 3 distinct peaks for FePc with the first peak position at \ang{6.93}. The small satellite peak, indicated with a dashed line, near the FePc(200) main peak is due to the reflection at the FePc/silicon interface and used to determine the FePc thickness. }
    \label{fig:xrd}
\end{figure}

For thin film preparation, iron phthalocyanine powder was commercially obtained from Sigma-Aldrich and further purified using a thermal gradient re-sublimation process. \added{The source material was loaded into a thermal evaporator and outgased before deposition at \SI{E-6}{mbar}.} During the deposition, the \added{carefully cleaned silicon} substrates are rotated and heated \added{before deposition} to control the FePc crystal size and surface morphology of the thin films.  A series of 7 FePc thin films are deposited via thermal evaporation onto silicon substrates by varying the deposition temperature from \SIrange{32}{230}{\degreeCelsius}, while maintaining the thickness at \SI{160}{nm} using a quartz crystal monitor. From Kiessig fringes in x-ray diffraction, we confirmed the absolute thickness and also the standing orientation of the crystals due to the presence of the $2\theta = $ \ang{6.93} peak \added{($a = $~\SI{2.55}{nm})} in the high-temperature samples, \added{shown in Fig.~\ref{fig:xrd}}. It corresponds to a $d$-lattice spacing of \SI{12.7}{\angstrom}, slightly smaller than the $d$-lattice spacings in low-temperature deposited thin films.\cite{miller_quantitative_2005} The x-ray data confirms that the molecule's $b$-axis is parallel to the surface and the growth of FePc on silicon is the same as on sapphire substrates.
 
\begin{figure} 
    \centering
    \includegraphics[width=\FIGW]{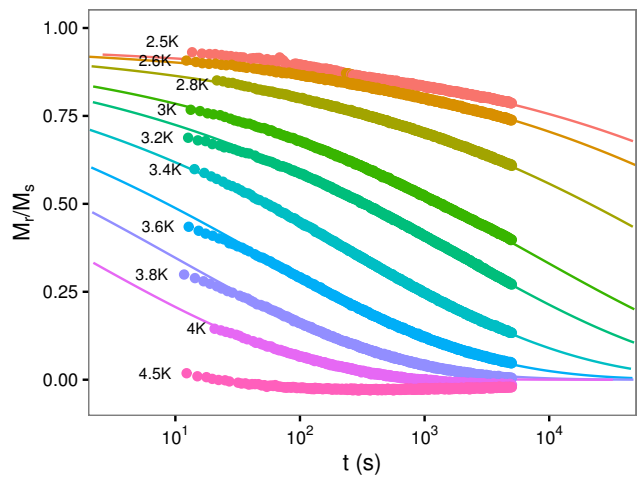}  
    \caption{The remanence $M_r$ for the FePc/Si sample deposited at \SI{180}{\degreeCelsius} shows slow relaxation at temperatures in the range of \SI{2.5}{K} and \SI{4.5}{K}. The lines are fits of the stretched exponential function from the master curve.}
    \label{fig:Temperatures}
\end{figure}

The magnetic properties of all thin films are measured in a commercial Quantum Design PPMS system using the vibrating sample magnetometry option at 40~Hz. The remanant state is created by cooling the sample from \SI{100}{K} in a magnetic field of 3~T to the desired measurement temperature. The field is always applied in-plane and then removed at a constant rate of \SI{0.01}{T/s} before the remanant magnetization is recorded as a function of time. Measurements last  at least \SI{5}{ks} at constant temperature.


The slow relaxation $M_r(t)$ shown in Fig.~\ref{fig:Temperatures} for a FePc thin film deposited at \SI{180}{\degreeCelsius} is representative for all samples. The maximum amount of relaxation within the measured time span occurs at intermediate temperatures near 3.3~K and depends on the deposition temperature. It shifts from \SI{2.7}{K} for short chains to \SI{3.3}{K} for longer chains. At and above \SI{4.5}{K} the remanent magnetization falls to near the detection limit, even though the saturation magnetization of the loop is not diminished much up to 15~K.

\begin{figure}  
   \centering
   \includegraphics[width=\FIGW]{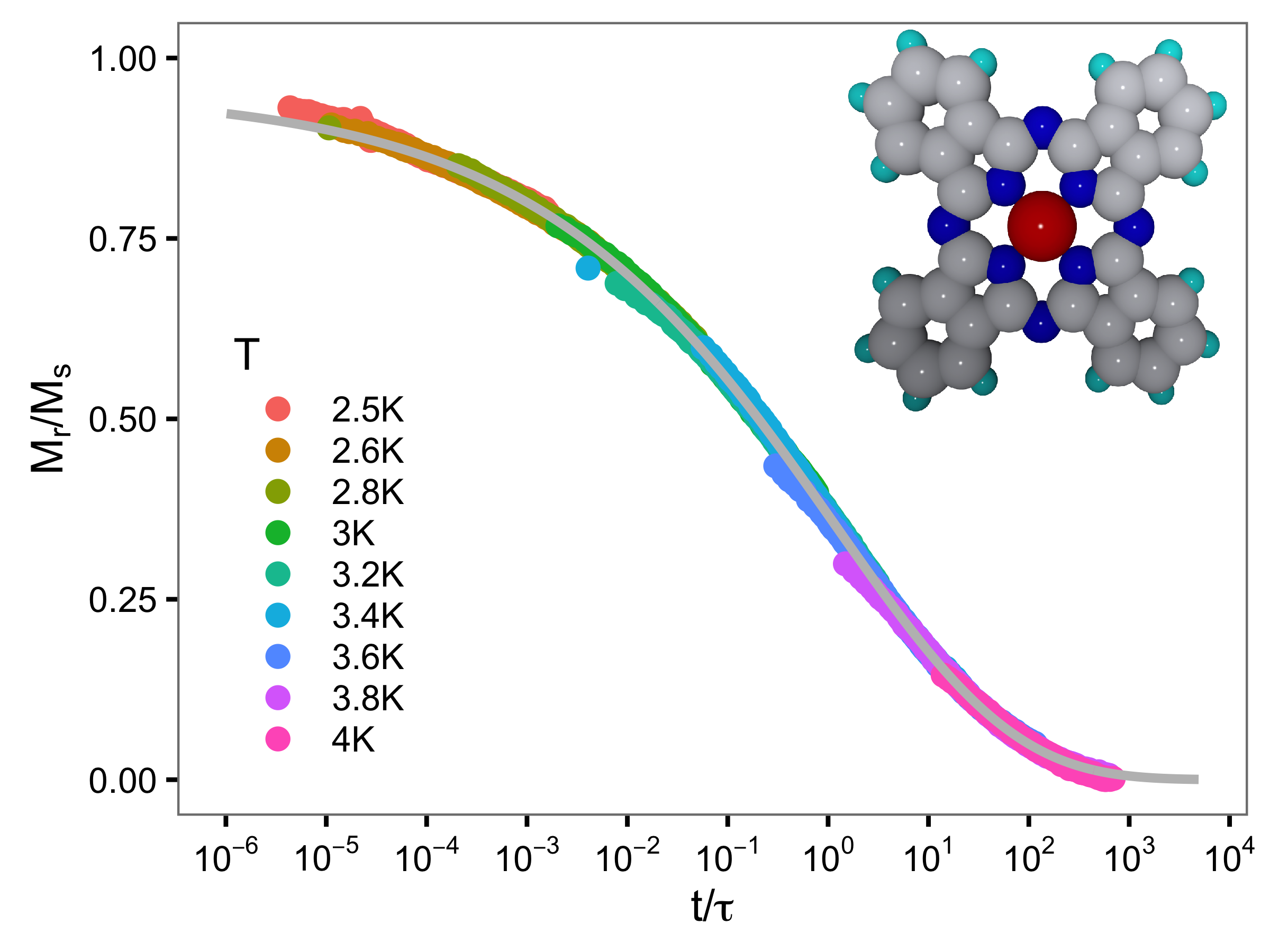}
   \caption{Scaled remanent measurements are collapsed onto a single master curve for temperatures from 2.5~K to 4~K for FePc/Si deposited at \SI{180}{\degreeCelsius}. Each measurement is scaled by the relaxation time $\tau$. The gray line shows a fit to the stretched exponential form with $\beta = 0.25$. Inset shows the FePc molecule with iron ion at the center.}
   \label{fig:mastercurve-predicted}
\end{figure}

Several models have been proposed to study systems that have distributions of relaxation times.\cite{zorn_logarithmic_2002}  Similar to relaxation in single chain magnets, we find empirically that the remanent magnetization $M_r(t, T)$ can be fit well to the following stretched exponential form

$$
M_{r} (t) =  M_0 \exp{ \left[ - (t/\tau)^{\beta}  \right] },
$$

with three fitting parameters, the relaxation time $\tau$, the amplitude $M_0$, and the stretch exponent $\beta$. At most measurement temperatures, the relaxation is very slow. Fitting the data from Fig.~\ref{fig:Temperatures} to  a stretched exponential, we find a non-zero offset and also non-convergent values of $\tau$. As the measurement time is increased from 5~ks to 10~ks, values for the relaxation time increase, and the offset decreases. A method to overcome the finite-sized measurement times in very slow relaxing systems is a scaling approach. Curves measured at different temperatures are scaled with their relaxation time to form a master curve, similar to work on Mn-Ni-Mn trimer single chain magnets.\cite{ferbinteanu_single-chain_2005} The relaxation time-scaled curve in Fig.~\ref{fig:mastercurve-predicted} spans more than 8 decades of time and is easily fit to the stretched exponential form without offset. In this method, we effectively use high-temperature relaxation data to extrapolate the slow relaxation measurements at low temperatures. The stretch exponent is fit to $\beta = 0.25$. It suggests very slow relaxation due to polydisperse samples. 

\begin{figure} 
   \centering
   \includegraphics[width=\FIGW]{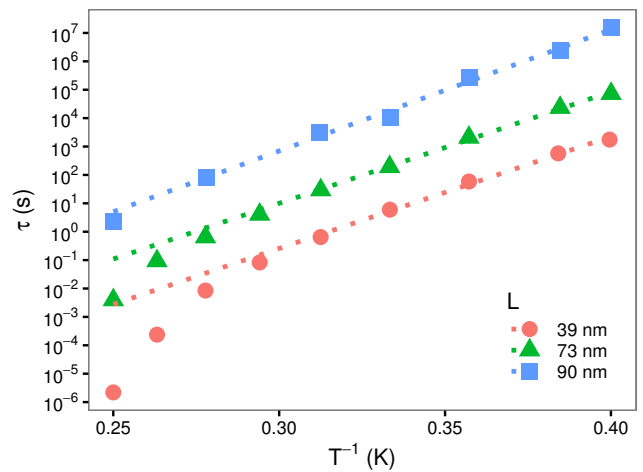}   
   \caption{The relaxation times are determined from fits to $M_r(t)$ using the stretched exponential form. Data for three samples are shown, the dotted lines are linear fits to extract the energy barrier $\Delta_A$ and the intercept $\tau_0$.}
   \label{fig:Arrhenius}
\end{figure}

\begin{figure}  
   \centering
   \includegraphics[width=\FIGW]{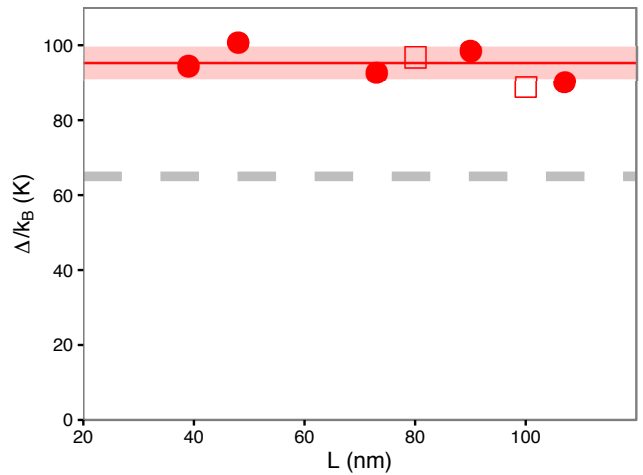}   
   \caption{Energy gap $\Delta_A/k_B$ versus chain length of sample. The gray line is added for reference for $\Delta_\xi/k_B = 64$~K determined from MCD data in FePc/Au thin films. The chain length of 5 samples (circle) is determined from the crystal size using AFM, the other two samples (square) are inferred lengths from coercivity measurements.}
   \label{fig:Delta}
\end{figure}

As observed for single chain magnets, the extracted relaxation time with temperature follows an Arrhenius behavior with a characteristic energy barrier height $\Delta_A$. In the form graphed in Fig.~\ref{fig:Arrhenius}, the slope represents the energy barrier height. The dashed lines correspond to linear fits and suggest that the slope is independent of the chain length $L$. In the Glauber-Ising model, the energy barrier is interpreted as the energy needed to flip a spin.  Applying fits to the low-temperature data ($T < 3.4$~K), the barrier height is determined as $\Delta_A/k_B = 95 (\pm4)$~K, see Fig.~\ref{fig:Delta}.  For single chain magnets, this barrier height is proportional to the anisotropy energy $D$.\cite{coulon_glauber_2004} It remains constant for all FePc samples. \added{For samples deposited under \SI{200}{\degreeCelsius}, the AFM crystal size and coercivity increase monotonically. For the highest deposition temperatures, the coercivity decreases, even though  AFM images show the longest crystals. It  suggests that the crystals are limited by defects.} The chain length in two samples deposited at high temperatures are estimated from coercivity data \added{instead of AFM images}. This results in nominal chain lengths between \SI{80}{nm} and \SI{100}{nm}. The data points are added to the graph in Fig.~\ref{fig:Delta} with open symbols to distinguish the low-temperature samples, for which the grain size is determined via atomic force microscopy (AFM).

As a reference, we estimate the correlation length and the associated energy barrier. Using data from magnetic circular dichroism measurements on FePc/Au/Si samples,\cite{gredig_substrate-controlled_2012} we can estimate  $\Delta_\xi$ using the relation that $\chi T \sim \exp (\Delta_{\xi}/k_B T)$ and fit the high-temperature data. We find $\Delta_\xi = 64$~K, which allows us to estimate the correlation length $\xi$ given the form $2 \xi / a = \exp(\Delta_\xi/k_B T)$.\cite{coulon_single-chain_2006} We conclude that $\xi \gg L$ for all measurements below \SI{5}{K}.
 \added{Given the different geometry of FePc/Si thin films, this solely serves as an estimate. } 

\begin{figure}  
   \centering
   \includegraphics[width=\FIGW]{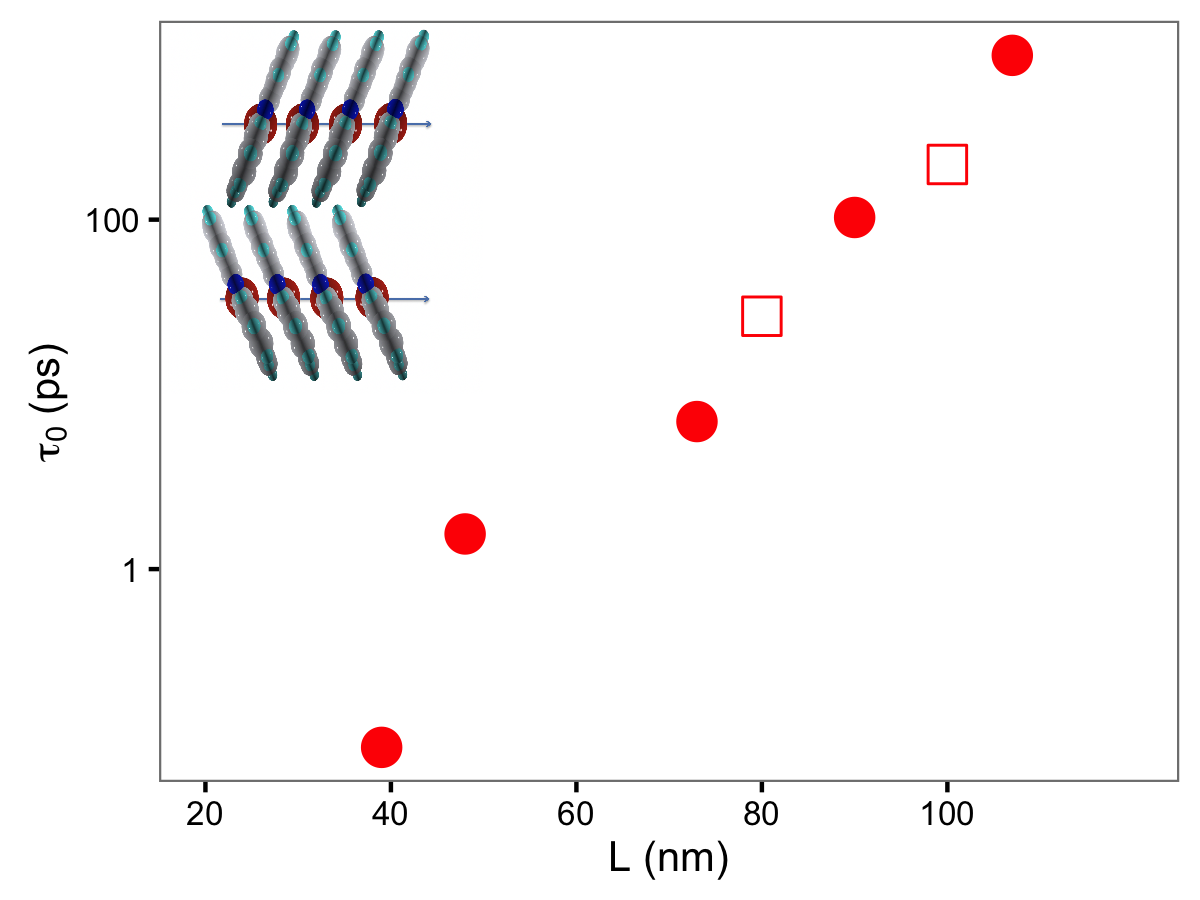}  
   \caption{Single spin relaxation times for 7 samples with different chain lengths. For two samples (squares) deposited above \SI{200}{\degreeCelsius} the length $L$ is estimated indirectly. Inset shows a top view of the herringbone arrangement of FePc molecules, the arrow indicates the direction of the magnetic field and the direction of the $b$-axis.
   }
   \label{fig:tau0}
\end{figure}

In the Glauber-Ising model, the single spin relaxation time $\tau_0$ is an adjustable parameter with a dependence on the chain length, if $\xi \gg L$. Experimentally, the spin relaxation time corresponds to the y-intercept in  Fig.~\ref{fig:Arrhenius} and varies with chain length. Small variations in the slope, however, amplify the intercept position. Therefore, the measured single-spin relaxation times $\tau_0$ should be interpreted with caution. We find a strong dependence of $\tau_0$ with the chain length $L$ as summarized in the graph of Fig.~\ref{fig:tau0}. The results are consistent with the impurity-doped crystals that show diminished values for $\tau_0$ as impurities are introduced and the findings agree with theoretical results of finite-sized magnetic chains that argue the single spin relaxation time is correlated with chain length, if the condition $\xi \gg L$ is met.\cite{leal_da_silva_critical_1995,luscombe_finite-size_1996,vindigni_finite_2004} Still, finite-size Ising chains are predicted to have a linear dependence on $L$, slower than our observations suggesting that the FePc thin film system may not fulfill all assumptions in the theoretical framework. \added{Indeed, the sub-lattice magnetization of a herringbone structure as shown in the inset of Fig.~\ref{fig:tau0} with an xy anisotropy in the plane of the molecule may result in more complex magnetic dynamics. This zig-zag structure essentially cancels one component, so that the net magnetization points along the $b$-axis, which is averaged over the plane.}

In summary, low-dimensional magnetic spin systems play a key role as paradigms for understanding dynamic magnetic behavior. The Glauber-Ising model has been extended to capture dynamic effects of finite-sized systems theoretically,\cite{leal_da_silva_critical_1995,luscombe_finite-size_1996,vindigni_finite_2004} but experimental realizations in powder and crystal samples suffer from random defects that prevent studying the explicit chain length dependence. Here, we propose to build nano-structures via the control of thin film deposition parameters to vary the chain length systematically.  The relaxation time of FePc thin films is extracted using dc measurements and shows strong dependence on the chain length $L$ in the regime of $\xi \gg L$.  The energy barrier $\Delta_A$ is independent of deposition temperature and chain length.  Using a simple molecule, iron phthalocyanine, we demonstrate the effect of tuning the chain length to achieve control of the magnetic relaxation time. Using templated substrates and careful growth conditions, the chain lengths in FePc thin films can be designed to achieve locally variable magnetic relaxation times. The system also provides a useful experimental realization of tunable finite-sized chain systems and experimental insight into magnetic dynamics of finite-sized ion chains, which are broadly accessible with theoretical tools.

\begin{acknowledgements}
This research was supported in part by the National Science Foundation (NSF) under grant NSF DMR~0847552. Insightful discussions with Dr.~Alessandro Vindigni are gratefully acknowledged. 
\end{acknowledgements}

\bibliography{/Users/Thomas/Documents/Papers/GredigLibrary}

\end{document}